\documentclass[titlepage,12pt]{article}
\usepackage{amssymb}
\usepackage{amsfonts}
\usepackage{amsmath}
\usepackage[font=small,format=plain,labelfont=bf,up,textfont=it,up]{caption}

\begin{document}

\title{\textbf{Missing solution in a Cornell potential}\thanks{%
To appear in Annals of Physics}}
\author{L.B. Castro\thanks{%
luis.castro@pgfsc.ufsc.br } \\
Departamento de F\'{\i}sica, CFM\\
Universidade Federal de Santa Catarina,\\
88040-900, Florian\'{o}polis - SC, Brazil\\\\
and\\\\
A.S. de Castro\thanks{%
castro@pesquisador.cnpq.br}\\
Departamento de F\'{\i}sica e Qu\'{\i}mica, Campus de Guaratin\-gue\-t\'{a}\\
Universidade Estadual Paulista,\\
12516-410, Guaratinguet\'{a} - SP, Brazil}
\date{}
\maketitle

\begin{abstract}
Missing bound-state solutions for fermions in the background of a Cornell
potential consisting of a mixed scalar-vector-pseudoscalar coupling is
examined. Charge-conjugation operation, degeneracy and localization are
discussed.

\bigskip


\bigskip


\bigskip \noindent Highlights:

\begin{itemize}
\item The Dirac equation with scalar-vector-pseudoscalar Cornell potential
is investigated. $\>$

\item The isolated solution from the Sturm-Liouville problem is found.

\item Charge-conjugation operation, degeneracy and localization are
discussed.
\end{itemize}
\end{abstract}

In a recent article, Hamzavi and Rajabi \cite{hr} investigated the Dirac
equation in a 1+1 dimension with the Cornell potential. They mapped the
Dirac equation into a Sturm-Liouville problem of a Schr\"{o}dinger-like
equation and obtained a set of bound-state solutions by recurring to the
properties of the biconfluent Heun equation. Nevertheless, an isolated
solution from the Sturm-Liouville scheme was not taken into account. The
purpose of this work is to report on this missing bound-state solution.

In the presence of time-independent interactions the most general 1+1
dimensional time-independent Dirac equation for a fermion of rest mass $m$
and momentum $p$ reads%
\begin{equation}
\left( c\sigma _{1}\hat{p}+\sigma _{3}m+\frac{I+\sigma _{3}}{2}V_{\Sigma }+%
\frac{I-\sigma _{3}}{2}V_{\Delta }+\sigma _{2}V_{p}\right) \psi =E\psi
\label{dirac1}
\end{equation}%
where $E$ is the energy of the fermion, $V_{\Sigma }=V_{v}+V_{s}$ and $%
V_{\Delta }=V_{v}-V_{s}$. The subscripts for the terms of potential denote
their properties under a Lorentz transformation: $v$ for the time component
of the two-vector potential, $s$ for the scalar potential and $p$ for the
pseudoscalar potential. $\sigma _{i}$ are the Pauli matrices and $I$ is the 2%
$\times $2 unit matrix. From now on we restrict our attention to the case $%
V_{\Delta }=0$. On that ground, Eq. (\ref{dirac1}) can also be written as
\begin{eqnarray}
-i\frac{d\psi _{-}}{dx}+m\psi _{+}+V_{\Sigma }\psi _{+}-iV_{p}\psi _{-}
&=&E\psi _{+}  \notag \\
&&  \label{dirac2} \\
-i\frac{d\psi _{+}}{dx}-m\psi _{-}+iV_{p}\psi _{+} &=&E\psi _{-}  \notag
\end{eqnarray}

\noindent where $\psi _{+}$ and $\psi _{-}$ are the upper and lower
components of the Dirac spinor respectively. It is clear from the pair of
coupled first-order differential equations (\ref{dirac2}) that both $\psi
_{+}$ and $\psi _{-}$ have opposite parities if the Dirac equation is
covariant under $x\rightarrow -x$. In this sense, $V_{p}$ changes sign
whereas $V_{\Sigma }$ remains the same. Because the invariance under
reflection through the origin, components of the Dirac spinor with
well-defined parities can be built. Thus, it suffices to concentrate
attention on the half line and impose boundary conditions on $\psi $ at the
origin and at infinity. Components of the Dirac spinor on the whole line
with well-defined parities can be constructed by taking symmetric and
antisymmetric linear combinations of $\psi $ defined on the half line:
\begin{equation}
\psi _{\pm }^{\left( \lambda \right) }\left( x\right) =\left[ \theta \left(
+x\right) \pm \lambda \,\theta \left( -x\right) \right] \psi _{\pm }\left(
|x|\right)
\end{equation}%
where $\lambda $ ($-\lambda $) denotes the parity of $\psi _{+}$ ($\psi _{-}$%
) and $\theta \left( x\right) $ the Heaviside step function. For a
normalized spinor, the expectation value of any observable $\mathcal{O}$ may
be given by
\begin{equation}
\left\langle \mathcal{O}\right\rangle =\int dx\,\psi ^{\dagger }\mathcal{O}%
\psi  \label{exp}
\end{equation}%
where the matrix $\mathcal{O\ }$must be Hermitian for insuring that $%
\left\langle \mathcal{O}\right\rangle $ is a real quantity. In particular, $%
\psi ^{\dagger }V\psi $ is an integrable quantity. Here $V$ is the potential
matrix%
\begin{equation}
V=\frac{I+\sigma _{3}}{2}\,V_{\Sigma }+\sigma _{2}V_{p}
\end{equation}

For $E\neq -m$, the searching for solutions can be formulated as a
Sturm-Liouville problem for the upper component of the Dirac spinor, as done
in Ref. \cite{hr} for bound states.

For $E=-m$, though, one can write
\begin{eqnarray}
\frac{d\psi _{+}}{dx}-V_{p}\psi _{+} &=&0  \notag \\
&&  \label{q1} \\
\frac{d\psi _{-}}{dx}+V_{p}\psi _{-} &=&-i\left( V_{\Sigma }+2m\right) \psi
_{+}  \notag
\end{eqnarray}%
whose solution is%
\begin{eqnarray}
\psi _{+}\left( x\right)  &=&N_{+}e^{+v\left( x\right) }  \notag \\
&&  \label{q2} \\
\psi _{-}\left( x\right)  &=&\left[ N_{-}-iN_{+}I\left( x\right) \right]
e^{-v\left( x\right) }  \notag
\end{eqnarray}%
where $N_{+}$ and $N_{-}$ are normalization constants, and%
\begin{eqnarray}
v\left( x\right)  &=&\int\nolimits^{x}dy\,V_{p}\left( y\right)   \notag \\
&&  \label{II} \\
I(x) &=&\int\nolimits^{x}dy\,\left[ V_{\Sigma }\left( y\right) +2m\right]
e^{+2v\left( y\right) }  \notag
\end{eqnarray}%
It is worthwhile to note that this sort of isolated solution cannot describe
scattering states and is subject to the normalization condition $\int
dx\,\left( |\psi _{+}|^{2}+|\psi _{-}|^{2}\right) =1$. Because $\psi _{+}$
and $\psi _{-}$ are normalizable functions, the possible isolated solution
presupposes $V_{p}\neq 0$. More precisely, the singularities of $V_{p}$ and
its asymptotic behaviour determines if the solution exists or does not exist
\cite{b1}-\cite{b3}.

For the mixed Cornell potential%
\begin{eqnarray}
V_{\Sigma } &=&-\frac{a_{\Sigma }}{|x|}+b_{\Sigma }|x|  \notag \\
&&  \label{pot} \\
V_{p} &=&-\frac{a_{p}}{x}+b_{p}x  \notag
\end{eqnarray}%
one finds $v\left( |x|\right) =-a_{p}\ln |x|+b_{p}x^{2}/2$ so that%
\begin{eqnarray}
\psi _{+}\left( |x|\right) &=&N_{+}|x|^{-a_{p}}e^{+b_{p}x^{2}/2}  \notag \\
&& \\
\psi _{-}\left( |x|\right) &=&\left[ N_{-}-iN_{+}I\left( |x|\right) \right]
|x|^{+a_{p}}e^{-b_{p}x^{2}/2}  \notag
\end{eqnarray}

A normalizable solution for $b_{p}>0$ is possible if $N_{+}=0$. Thus,
\begin{equation}
\psi \left( |x|\right) =N|x|^{a_{p}}e^{-|b_{p}|x^{2}/2}\left(
\begin{array}{c}
0 \\
1%
\end{array}%
\right) ,b_{p}>0  \label{isol1}
\end{equation}%
regardless of $a_{\Sigma }$, $b_{\Sigma }$ and $m$. Note that (\ref{isol1})
is square-integrable even if it is singular at the origin for $-1/2<a_{p}<0$%
. Nevertheless, near the origin
\begin{equation}
\hat{p}^{2}\psi \varpropto a_{p}\left( a_{p}-1\right) |x|^{a_{p}-2}
\end{equation}%
so that the operator $\hat{p}^{2}$ may not be a legitimate linear
transformation, because it may carry functions out of Hilbert space. This
observation allows us to impose additional restrictions on the
eigenfunction. $\hat{p}^{2}\psi $ is in Hilbert space only if
\begin{equation}
a_{p}=0\text{ or }a_{p}=1\text{ or }a_{p}>3/2
\end{equation}

As for $b_{p}<0$, a normalizable solution requires $N_{-}=0$, and a good
behaviour of $\psi _{+}$ near the origin, in the sense of normalization,
demands $a_{p}<1/2$. We must, however, calculate $I(|x|)$ to make explicit
the behaviour of $\psi _{-}$. For the Cornell potential given by (\ref{pot}%
), $I(|x|)$ can be expressed in terms of the incomplete gamma function \cite%
{abr}%
\begin{equation}
\gamma \left( a,z\right) =\int\nolimits_{0}^{z}dt\,e^{-t}t^{a-1},\text{ }%
\text{Re }a>0
\end{equation}%
As a matter of fact,
\begin{eqnarray}
I\left( |x|\right) &=&\frac{|b_{p}|^{a_{p}-1/2}}{2}\left[ 2m\gamma \left(
-a_{p}+1/2,|b_{p}|x^{2}\right) \right.  \notag \\
&&\left. +\frac{b_{\Sigma }}{\sqrt{|b_{p}|}}\gamma \left(
-a_{p}+1,|b_{p}|x^{2}\right) -a_{\Sigma }\sqrt{|b_{p}|}\gamma \left(
-a_{p},|b_{p}|x^{2}\right) \right]
\end{eqnarray}%
It follows immediately from the condition imposed on the definition of the
incomplete gamma function that $a_{p}<1/2$ if $m\neq 0$, $a_{p}<1$ if $%
b_{\Sigma }\neq 0$ and $a_{p}<0$ if $a_{\Sigma }\neq 0$. Now, because $%
\gamma \left( a,z\right) $ tends to $\Gamma \left( a\right) $ as $z$ tends
to infinity, $\psi _{-}$ is not, in general, a square-integrable function.
An exception, though, occurs when $m=a_{\Sigma }=b_{\Sigma }=0$ ($I(|x|)=0$)
just for the reason that $\psi _{-}$ vanishes identically. Therefore,%
\begin{equation}
\psi \left( |x|\right) =N|x|^{-a_{p}}e^{-|b_{p}|x^{2}/2}\left(
\begin{array}{c}
1 \\
0%
\end{array}%
\right) ,\text{ }b_{p}<0,\text{ }m=a_{\Sigma }=b_{\Sigma }=0  \label{isol2}
\end{equation}%
In this case, near the origin
\begin{equation}
\hat{p}^{2}\psi \varpropto a_{p}\left( a_{p}+1\right) |x|^{-a_{p}-2}
\end{equation}%
so that $\hat{p}^{2}\psi $ is in Hilbert space only if
\begin{equation}
a_{p}=0\text{ or }a_{p}=-1\text{ or }a_{p}<-3/2
\end{equation}

It is instructive to note that
\begin{equation}
\psi ^{\dagger }V\psi =V_{\Sigma }|\psi _{+}|^{2}-2V_{p}\,\text{Im }\psi
_{-}^{\ast }\psi _{+}
\end{equation}%
so that all the isolated solutions make $\left\langle V\right\rangle =0$.
Notice from (\ref{q1}) that when $m=0$ and $V_{\Sigma }=0$, $\psi _{+}$
turns into $\psi _{-}$ and vice versa as $V_{p}$ changes its sign. As a
matter of fact, these zero-energy solutions are related by the
charge-conjugate operation \cite{asc}.

The extension of $\psi $ compatible with (\ref{q1}) for the whole line
demands that $\psi _{+}$ and $\psi _{-}$ be continuous at the origin,
otherwise (\ref{q1}) would contain the $\delta $-functions. On the whole
line they are even functions if $a_{p}=0$. If $|a_{p}|=1$ or $|a_{p}|\geq 3/2
$, though, the eigenvalue $E=-m$ is two-fold degenerate due to the existence
of symmetric and asymmetric extensions of $\psi $. It is worthwhile to note
that this sort of degeneracy for bound-state solutions in a one-dimensional
system contrasts with a well-known nondegeneracy theorem in the
nonrelativistic theory \cite{lan}. However, it has been shown that the
theorem is not necessarily valid for singular potentials \cite{lou}. Our
results are summarized in Table 1.

\begin{table}[th]
\begin{center}
\begin{tabular}{|c|}
\hline
\\
$\psi \left( |x|\right) =\mathcal{S}N|x|^{|a_{p}|}e^{-|b_{p}|x^{2}/2}$ \\
\\
sgn$(b_{p})=$sgn$(a_{p}),\,|a_{p}|=0\;$or $1\,$or$\;>3/2$ \\
\\ \hline
\\
only symmetrical extension for $a_{p}=0$ \\
\\
\\
symmetrical and antisymmetrical extensions for $|a_{p}|=1\;$or$\;|a_{p}|>3/2$
\\
\\ \hline
\\
$\mathcal{S}=\left\{
\begin{array}{cc}
\left(
\begin{array}{c}
0 \\
1%
\end{array}%
\right) , & \;\text{for}\;b_{p}>0\hspace{1.65in} \\
&  \\
\left(
\begin{array}{c}
1 \\
0%
\end{array}%
\right) , & \text{for}\;b_{p}<0\;\text{with}\;m=a_{\Sigma }=b_{\Sigma }=0%
\end{array}%
\right. $ \\
\\ \hline
\end{tabular}%
\end{center}
\caption{Summary of the results for the solution with $E=-m$ ($\left\langle
V\right\rangle =0$) defined on the whole line. sgn($z$) stands for the sign
function of $z$. For massless fermions, the zero energy solutions with $%
b_{p}>0$ and $b_{p}<0$ are related by charge-conjugate operation when $%
a_{\Sigma }=b_{\Sigma }=0$.}
\end{table}

The uncertainties in the position and momentum for the solution on the whole
line can be written as

\begin{equation}
\Delta x=\frac{1}{\sqrt{|b_{p}|}}\sqrt{|a_{p}|+1/2}
\end{equation}%
\begin{equation}
\Delta p=\sqrt{|b_{p}|}\sqrt{\frac{|a_{p}|-1/4}{|a_{p}|-1/2}}
\end{equation}%
If $\Delta x$ shrinks then $\Delta p$ will must swell, in consonance with
the Heisenberg uncertainty principle. Notice that $\Delta x\Delta p$ is
independent of $|b_{p}|$, and when $a_{p}=0$ one obtains the
minimum-uncertainty solution ($\Delta x\Delta p=1/2$) due to the Gaussian
form of the eigenspinor, as it happens in the ground-state solution for the
nonrelativistic harmonic oscillator. Nevertheless, the maximum uncertainty
in the momentum is comparable with $m$ requiring that is impossible to
localize a fermion in a region of space less than or comparable with half of
its Compton wavelength. This impasse can be broken by resorting to the
concepts of effective mass and effective Compton wavelength \cite{b1}, \cite%
{b2}. Indeed, if one defines an effective mass as $m_{\mathtt{eff}}=\sqrt{%
|b_{p}|}$ and an effective Compton wavelength as $\lambda _{\mathtt{eff}%
}=1/m_{\mathtt{eff}}$ one will find that the high localization of fermions,
related to high values of $|b_{p}|$ never menaces the single-particle
interpretation of the Dirac theory. In fact, $\left( \Delta x\right) _{\min
}=\sqrt{2}\lambda _{\mathtt{eff}}/2$ occurs for $a_{p}=0$, and $\left(
\Delta p\right) _{\max }=\sqrt{6}m_{\mathtt{eff}}/2$ for $a_{p}=1$.

\bigskip

\bigskip

\bigskip

\bigskip

\bigskip

\bigskip

\bigskip

\bigskip

\noindent \textbf{Acknowledgments}

This work was supported in part by means of funds provided by CAPES and
CNPq. This work was partially done during a visit (L.B. Castro) to
UNESP-Campus de Guaratin\-gue\-t\'{a}.

\end{document}